# Metamaterial slab as a lens, a cloak or something in between

Jian Wen Dong[1,2,*], Hui Huo Zheng[2,*], Yun Lai[2], He Zhou Wang[1] and C. T. Chan[2,†]

[1.] State Key Laboratory of Optoelectronic Materials and Technologies, Sun Yat-Sen (Zhongshan) University, Guangzhou 510275, China

[2.] Department of Physics, The Hong Kong University of Science and Technology, Hong Kong, China

* Equal contributions.

† Corresponding author. Email: phchan@ust.hk

We show that a metamaterial slab with arbitrary values of $\varepsilon$ and $\mu$ behaves as a cloak at a finite frequency for a small object located sufficiently close to it due to the suppression of the object's optical excitations by enhanced reflections. Reflections due to propagating components can partially suppress the excitation while evanescent components can cloak the object completely. In particular, a Veselago slab with $\varepsilon = \mu = -1 + i\delta$, as well as a class of anisotropic negative refractive index slabs, can completely cloak the small object placed within a finite distance from the slab when $\delta \to 0$.

An ideal Veselago slab with $\varepsilon = \mu = -1$ [1,2] can focus all the Fourier components of light, making it a perfect lens. How it works can be explained with the concept of "complementary media" [3], which states that any medium can be optically cancelled by a material of equal thickness constructed as an inverted mirror image of the medium, with $(\varepsilon, \mu)$ reversed in sign. When an object placed at a distance $z_d$ from the front of the Veselago slab with $\varepsilon = \mu = -1$ and thickness $d$, a perfect image will be formed $2d - z_d$ away from the front of the slab. However, the perfect lens effect is subtle (see, for example, [4] for some early discussions on this). In particular, an infinitesimal loss is usually required in the Veselago slab, but it will also induce localized resonances which influence the image formation of an object lying within the resonance region [5-8]. As the loss approaches zero, any point object located less than a certain distance from the lens will be cloaked due to the presence of the resonant fields [7]. The $\varepsilon = \mu = -1$ slab is thus a perfect lens and also a perfect cloak, which is apparently contradictory. Our motivation is to solve this puzzle and go beyond the perfect lens ($\varepsilon = \mu = -1$) by considering a metamaterial slab with arbitrary values of $(\varepsilon, \mu)$.

Using a Green's function method, we find that for a slab with arbitrary values of $\varepsilon$ and $\mu$, the reflections from the slab (either the propagating components or the evanescent components) can suppress the excitation of a small object placed close to it. The evanescent components can completely suppress the excitation and cloak the object as $z_d \to 0$. This cloaking effect is generally not influenced by losses in the slab. But the lossy Veselago slab is a special case in which the suppression effect of the evanescent wave is so strong that it completely cloaks small objects within a finite distance of $d/2$ as the material loss goes to zero. We also find that a class of "folded geometry" slabs [9,10] has the same cloaking effects for a small object within a certain critical distance in the limit of zero absorption. A negative index metamaterial slab generally behaves as somewhere between a cloak and a lens in the sense that it will form an image with suppressed intensity for a small object placed close to it.

We consider a passive object that is small enough to be represented by a passive dipole and place it in front of a metamaterial slab, and we illuminate the system using an external light source. The object has a dynamic dipole polarizability, $\alpha = i(3/2k_0^3)a_1$, where $k_0 = \omega/c$ with $c$ as the speed of light, and $a_1$ is the electric term of the Mie's coefficients [11]. The slab is placed in the *xy* plane and the external light source is a dipole source placed vertically above the passive object with a polarization along the y direction. The induced dipole moment on the object is $\mathbf{p}_0 = \alpha\left(\mathbf{E}_s^{ext} + \mathbf{E}_0^{ref}\right)$, where $\mathbf{E}_s^{ext} = 4\pi k_0^2 \mathbf{W}^{tot} \cdot \mathbf{p}_{src}$ is the external field due to the dipole source in the presence of the slab, and $\mathbf{E}_0^{ref} = 4\pi k_0^2 \mathbf{W}^{ref} \cdot \mathbf{p}_0$ is the reflected field from the slab due to the passive dipole itself. Here, $\mathbf{p}_{src}$ is the external dipole source. $\mathbf{W}^{tot}$ is the dyadic Green's function which takes into account the effect of the slab, and $\mathbf{W}^{ref}$ is the reflection part of the dyadic Green's function. Since the external dipole source is polarized along the y direction, and both the active dipole and the passive object are placed on the z-axis, the nonzero component of the induced dipole moment is also oriented in the y direction,

$$p_y = 4\pi k_0^2 \frac{W_{yy}^{tot}}{\alpha^{-1} - 4\pi k_0^2 W_{yy}^{ref}} p_{src}. \tag{1}$$

We can define an effective polarizability along the y-direction,

$$\alpha^* = \left(\alpha^{-1} - 4\pi k_0^2 W_{yy}^{ref}\right)^{-1}, \tag{2}$$

which is the key to understand the functionality of the slab. We note that this source dipole configuration is arranged to excite a dipole parallel to the slab through one principal component ($\alpha^* = \alpha_{yy}^*$) of the effective polarizability tensor. By symmetry, $\alpha_{xx}^* = \alpha_{yy}^*$. The other principal value of effective polarizability tensor, $\alpha_{zz}^*$, which corresponds to the excited dipole being perpendicular to the slab, will be discussed in the Auxiliary Material (Part B), and the results are qualitatively the same.

As the field at the image point is $\mathbf{E}(\mathbf{r}_i) = \alpha^* \mathbf{W}^{tot}(\mathbf{r}_i, \mathbf{r}_o) \cdot \mathbf{E}_s^{ext}(\mathbf{r}_o)$, with $\mathbf{r}_o$ and $\mathbf{r}_i$ being the object position and image position respectively, the slab serves as a cloak if $\alpha^* \to 0$. The slab is potentially a lens if $\alpha^* \to \alpha$. The quality of the image depends on the material parameters. From Eq. (2), we find $\alpha^* \to 0$ if $W_{yy}^{ref}$ diverges, while $\alpha^* \to \alpha$ if $W_{yy}^{ref} \to 0$. The explicit expressions for the Green's functions can be found in the literature [12]. In particular, $W_{yy}^{ref}$ has the form:

$$W_{yy}^{ref} = \frac{i}{8\pi} \int_0^\infty \frac{k_{//}}{k_{0z}} dk_{//} \left( R^{TE} e^{i2k_{0z}z_d} - \frac{k_{0z}^2}{k_0^2} R^{TM} e^{i2k_{0z}z_d} \right), \tag{3}$$

where $R^{TE}$ and $R^{TM}$ are the reflection coefficients of the slab, with the form of

$$R = \frac{2i(\zeta^2-1)\sin k_z d}{(\zeta+1)^2 e^{-ik_z d}-(\zeta-1)^2 e^{ik_z d}}, \text{ where } \zeta^{TE}=\frac{k_z}{k_{0z}\mu} \text{ and } \zeta^{TM}=\frac{k_z}{k_{0z}\varepsilon}, \text{ and } k_z^2+k_{//}^2=\varepsilon\mu k_0^2$$

[12].

We first consider the general cases in which $\varepsilon$ and $\mu$ are not both -1. Figures 1(a) and 1(b) show the computed values of $|\alpha^*/\alpha|$ with respect to the distance $z_d$ for two cases, namely $\varepsilon=-2+i\delta, \mu=-3+i\delta$ and $\varepsilon=\mu=2+i\delta$, with a few values of $\delta$. We see that as $z_d \to 0$, the slab will suppress the dipole excitation (i.e., $|\alpha^*/\alpha| \to 0$). This suppressing phenomenon is found to be universal for any value of $(\varepsilon,\mu)$, as is seen from the asymptotic behavior of Eq. (3) for small $z_d$,

$$W_{yy}^{ref} \to \frac{1}{16\pi}\left[R_{\lim}^{TM}\left(\frac{\kappa_0^2}{k_0^2 z_d}+\frac{\kappa_0}{k_0^2 z_d^2}+\frac{1}{2k_0^2 z_d^3}\right)+\frac{R_{\lim}^{TE}}{z_d}\right]e^{-2\kappa_0 z_d}, \tag{4}$$

where $\kappa_0 = 10\max\left\{\frac{1}{d}\log\left|\frac{\mu-1}{\mu+1}\right|,\frac{1}{d}\log\left|\frac{\varepsilon-1}{\varepsilon+1}\right|,\sqrt{\varepsilon\mu}k_0\right\}$, $R_{\lim}^{TE}=\frac{\mu-1}{\mu+1}$, and $R_{\lim}^{TM}=\frac{\varepsilon-1}{\varepsilon+1}$ [13].

Eq. (4) indicates that $4\pi k_0^2 W_{yy}^{ref}$ diverges as $z_d^{-3}$ as $z_d$ approaches zero. This universal result shows that the reflection of the evanescent wave in the slab with sufficiently large Fourier component, $\kappa_0$, can suppress the dipole excitation ($\alpha^* \to 0$) completely as long as the object is sufficiently close to the slab. This suppression effect occurs for any material parameter value (except for when $\varepsilon=\mu=1$), but the effect is stronger in negative-refractive-index slabs than in positive-refractive-index slabs (with the same absolute values of $\varepsilon,\mu$) due to larger values of $R_{\lim}^{TE}$ and $R_{\lim}^{TM}$.

The suppression effect is strongest as $\varepsilon, \mu \to -1$, which is a case requiring special treatment. We consider a finite value of the absorption parameter $\delta$, and consider in the limit $\delta \to 0$. As shown in Fig. 1(c), there is a critical distance of $d/2$ at which $\alpha^*$ drops to zero abruptly when $\delta \to 0$. It indicates that the Veselago slab behaves as a cloak with a "suppression zone" of thickness $d/2$ as long as the absorption is vanishingly small. This is the finite frequency analog of the "anomalous resonance effect" first discovered by Nicorovici, McPhedran, and Milton [5]. Such a cloaking effect can be analytically deduced from the asymptotic behavior of $W_{yy}^{ref}$ for small $\delta$, which can be written as [13]

$$W_{yy}^{ref} \to \frac{i}{4\pi\delta^\gamma d}\left[I_0(\gamma)\left(1+\frac{\log^2\delta}{k_0^2 d^2}\right) - 2\frac{\log\delta}{k_0^2 d^2}I_1(\gamma) + \frac{1}{k_0^2 d^2}I_2(\gamma)\right], \qquad (5)$$

where $\gamma = 1 - 2z_d/d$, and $I_n(\gamma) = \int_0^{+\infty} \frac{x^\gamma (\log x)^n}{x^2+4}dx$ is a constant. Eq. (5) indicates that $4\pi k_0^2 W_{yy}^{ref}$ diverges if $z_d < d/2$ whereas $W_{yy}^{ref} \to 0$ if $z_d > d/2$ in the limit of zero absorption. Figures 1(a) and 1(b) show that the suppression effect depends only weakly on $\delta$ for general values of $\varepsilon, \mu$. Figure 1(c) shows that the case $\mathrm{Re}(\varepsilon) = \mathrm{Re}(\mu) = -1$ is special in the sense that the suppression effect is strongly dependent on $\delta$ and a finite "suppression zone" of $z_d = d/2$ emerges in the limit of small absorption [7].

When either $\varepsilon$ or $\mu$ equals to -1, the suppression of excitation is stronger than that when neither $\varepsilon$ nor $\mu$ equals to -1, but not as strong as that when both are equal to -1. Specifically, it can be shown [13] that for $\varepsilon = -1 + i\delta$, $\mu \neq -1$, $\alpha^*$ smoothly approaches zero with an asymptotic form of $z_d^5$ at a finite frequency. There is no critical distance within which the excitation suddenly turns zero in the limit $\delta \to 0$, but the suppression is stronger than that with any other values of $\varepsilon$. However, in the quasi-static limit, and more specifically [13] for $k_0 < \sqrt{2}\dfrac{\delta^{1/2}\log(2/\delta)}{d\sqrt{1+\mu}}$, there is a "suppression zone" of $d/2$ in the limit $\delta \to 0$. Besides, for $\mu = -1+i\delta$, $\varepsilon \neq -1$, $\alpha^*$ smoothly approaches zero with an asymptotic form of $z_d^3$ so that the suppression is not as strong as that in the case of $\varepsilon = -1+i\delta, \mu \neq -1$.

To further illustrate the complex behavior of $\alpha^*$ at a finite distance of $z_d$ for different material parameters ( $\varepsilon = \mu = x+i\delta$ ), we compute $\alpha^*$ numerically for various combinations of $(x,\delta)$, as shown in Fig. 2. The object is placed at a distance $z_d = d/5$ from the slab. We see that the color coding is mostly "yellow" which means that $|\alpha^*/\alpha|$ is approximately equal to 1 in most of the cases, so that the metamaterial slab behaves more or less as a lens. But for some particular values of $(x,\delta)$ in which the color is "blue", the value of $|\alpha^*/\alpha|$ is small and the slab behaves as a cloak. Figure 3 plots the relationship between $|\alpha^*/\alpha|$ and $x$ at different values of $z_d$ for a fixed small loss of $\delta = 10^{-5}$. We can see that $|\alpha^*/\alpha| \sim 1$ no matter what the value of $x$ is if the dipole object is far from the slab (the blue curve), showing the absence of cloaking. This is true for all $z_d > d/2$. However, if the object is moved very close to the slab, $|\alpha^*/\alpha|$ decreases significantly for any value of $x$ (the red curve), indicating the suppression of the dipole excitation by the metamaterial slab. While the suppression is strongest at $x = -1$, there are other values of $x$ that would give fairly strong suppression (see the dips in the black and red curves).

A careful analysis shows that there are two distinct mechanisms behind the suppression of $\alpha^*$. We note from Eq. (3) that both the propagating and evanescent components contribute to reflections. When the dipole is relatively far away from the slab, since the evanescent components ($k_{//} > k_0$) decay spatially (due to the $e^{i2k_{0z}z_d}$ factor in Eq. (3)), the suppression is mainly attributed to the propagating components ($k_{//} < k_0$). We find that the reflection coefficient, $R$, takes its optimal values when $\sin k_z d \sim \pm 1$, i.e., $\sqrt{\varepsilon\mu}k_0 d \sim (2n+1)\pi/2$, where $n$ is an integer. This causes the Fabry-Perot resonance of $\alpha^*$ at some particular values of $x \sim (\lambda/d)(2n+1)/4$, in the limit of large $|x|$. We note that for the relatively large values of $z_d$, large phase differences ($e^{i2k_{0z}z_d}$) will cause different propagating components to cancel each other. This explains why the resonant behaviors are stronger on the black and red curves than on the blue curve.

While the Fabry-Perot resonance can cause suppression, it cannot bring $|\alpha^*/\alpha|$ to exactly zero. For example, we see from Fig. 4(a) that at $\text{Re}(\varepsilon) = \text{Re}(\mu) = x = -1.953132$, $|\alpha^*/\alpha|$ reaches a very small value, but it remains finite for any value of absorption. This is because $R$ for $k_{//} < \sqrt{\varepsilon\mu}k_0$ is always of the order of unity. This implies that the suppression effect due to the propagating waves can only give a partial suppression. The complete cloaking effect is due to the strong reflection of the evanescent waves. When the object is moved closer to the slab, the large evanescent components ($k_{//} > \sqrt{\varepsilon\mu}k_0$) are expected to become more dominant, leading to stronger suppression of $\alpha^*$. One can see that $R$ approaches a constant $R_{\lim}^{TE}$ for TE mode (and $R_{\lim}^{TM}$ for TM mode) for $k_{//} > \frac{1}{d}\log\left|\frac{x-1}{x+1}\right|$. It yields $\left|\frac{\alpha^*}{\alpha}\right| \sim C\left|\frac{x-1}{x+1}\right|$, where $C$ is a constant that becomes smaller as $z_d$ decreases. This is exactly the profile of the red curve in Fig. 3. The suppression effect is the strongest at $x = -1$ which corresponds to the Veselago slab. Figure 4(b) shows the numerical results near $x = -1$ at finite $z_d$. It can be seen that the value of $|\alpha^*/\alpha|$ at $x = -1$ is much smaller than those near $x = -1$ as the loss goes to zero [14].

The next question is whether the Veselago slab is unique in the sense that it has a finite size "suppression zone". We find that suppression zones can also occur in other negative index slabs if we allow anisotropy. From the point of view of transformation optics [15], negative refractive index material parameters correspond to a "folded geometry" coordinate transformation [9,10] with a negative slope which maps a single point in the electromagnetic space ($z'$) to multiple points in the physical space ($z$). This kind of mapping corresponds to $\varepsilon = \mu = diag\left(\frac{dz'}{dz}, \frac{dz'}{dz}, \frac{dz}{dz'}\right)$ in the physical space. The mapping $dz'/dz = -1$ gives the perfect lens. In general, a slab with $\varepsilon = \mu = diag(-\beta, -\beta, -1/\beta)$, where $\beta > 0$, corresponds to the mapping $dz'/dz = -\beta$ inside the lens and $dz'/dz = 1$ outside the lens. The reflection coefficient $R$ for evanescent components is the same as that of a Veselago lens of $\beta d$ [13], therefore the cloaking effect also occurs in such slabs with a critical cloaking distance $z_d = \beta d / 2$, which can be tuned to be as large as we like. We demonstrate numerically that the cloaking effect in the anisotropic media can be extended spatially out to a distance of $z_d = \beta d / 2$ [13].

We also note that the critical distance cloaking effect is mainly attributed to large $k_{//}$ components, and occurs at $(x, \delta) \rightarrow (-1, 0)$. In experimentally realizable metamaterials, the absorption is not that small and there is always some deviation from the exact value of $x = -1$. As the building blocks of metamaterials are discrete, there is always a cut-off in the parallel component of the wavevector. Hence most perfect lens experiments should observe the imaging effect. But in the limit of an ideal $\varepsilon = \mu = -1$ Veselago slab with infinitesimal loss, the image for a small object should have zero brightness if it is placed within $d/2$ of the slab. We also note that our results do not contradict the concept of "complementary media" [3]. If there is absorption in the perfect lens (e.g. $\varepsilon = \mu = -1 + i\delta$), a layer of air with "anti-absorption" (i.e. $\varepsilon = \mu = 1 - i\delta$) is required to maintain symmetry, and the pair of $\varepsilon = \mu = [-1 + i\delta, 1 - i\delta]$ does function as a perfect lens for any finite value of $\delta$.

In summary, we have investigated the imaging/cloaking condition of a small object in front of a metamaterial slab with arbitrary values of $(\varepsilon, \mu)$. The dipole excitation of a small object can be suppressed if it is placed sufficiently close to the slab. The suppression is strongest for "folded geometry" slabs which have finite suppression zones of width $\beta d/2$ in the limit of zero absorption.

This work was supported by the Hong Kong RGC (grant no. 600209). JWD and HZW are also supported by the NSFC of China (grant nos. 10804131, 10874250, 10674183), the FRFCU grant (2009300003161450), and the GDNSF grant. We thank Prof. Z. F. Lin, Dr. J. Ng, X. Q. Huang, Z. H. Hang, and J. Mei for thoughtful discussions. CTC thanks Prof. G. Milton for stimulating communications which motivated this study. Computation resources were partially supported by the HPCCC at HK Baptist University.

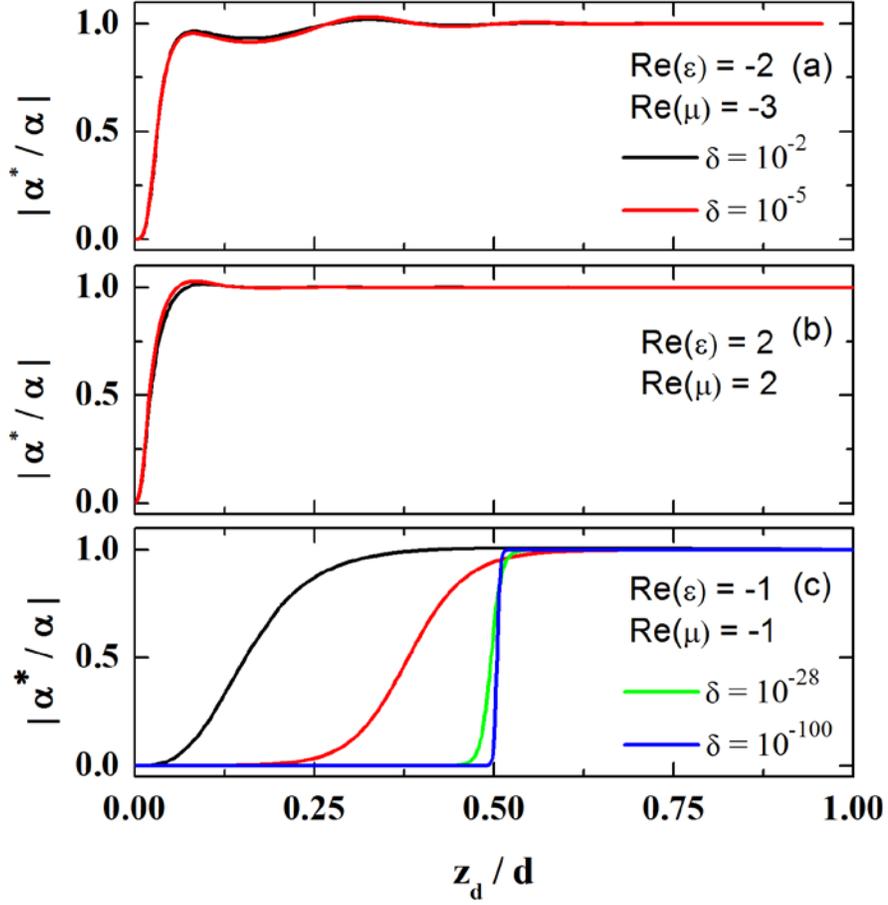

**FIG. 1.** (Color online) The suppression of effective polarizability (see text) as a function of $z_d/d$, where $d$ is the slab thickness and $z_d$ is distance between the dipole and the slab. $\delta$ is the loss parameter. The slab parameters are (a) $\mathrm{Re}(\varepsilon)=-2, \mathrm{Re}(\mu)=-3$; (b) $\mathrm{Re}(\varepsilon)=\mathrm{Re}(\mu)=2$; and (c) $\mathrm{Re}(\varepsilon)=\mathrm{Re}(\mu)=-1$, and in this special case a finite-size "suppression zone" of width $d/2$ emerges in the limit of $\delta \to 0$.

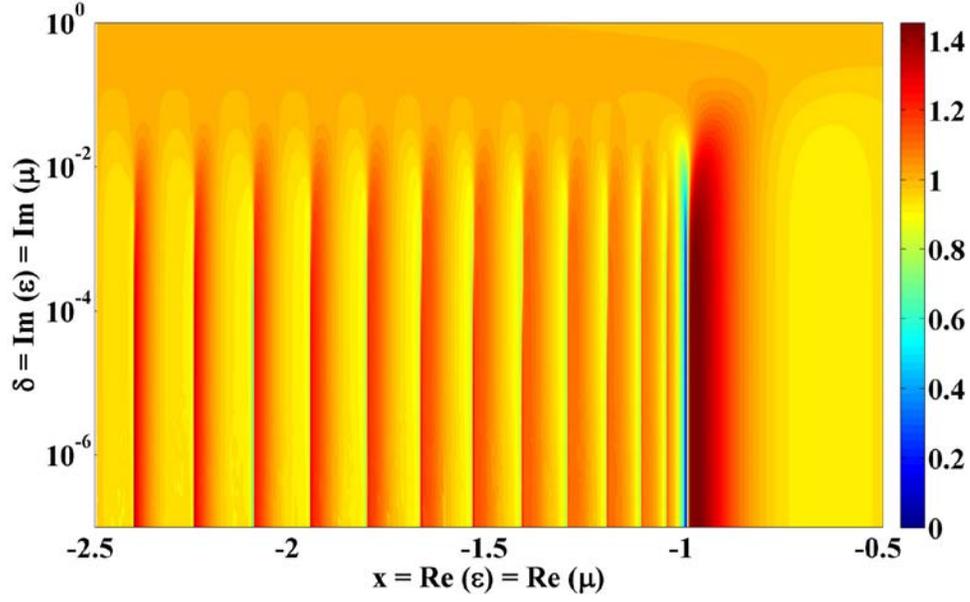

**FIG. 2. (Color online) The effective polarizability for a dipole positioned at** $z_d = d/5$ **in front of a slab with** $\varepsilon = \mu = x + i\delta$, **thickness** $d$ **and** $k_0 d = 6\pi$. **Blue color indicates strong suppression of excitation.**

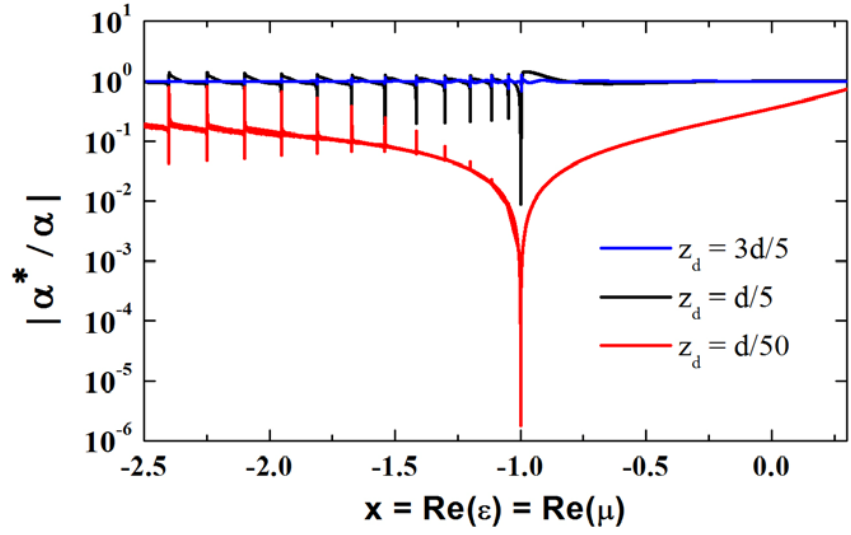

**FIG. 3. (Color online) The effective polarizability as a function of the real part of the permittivity of the slab with $\delta = 10^{-5}$ and different values of $z_d$.**

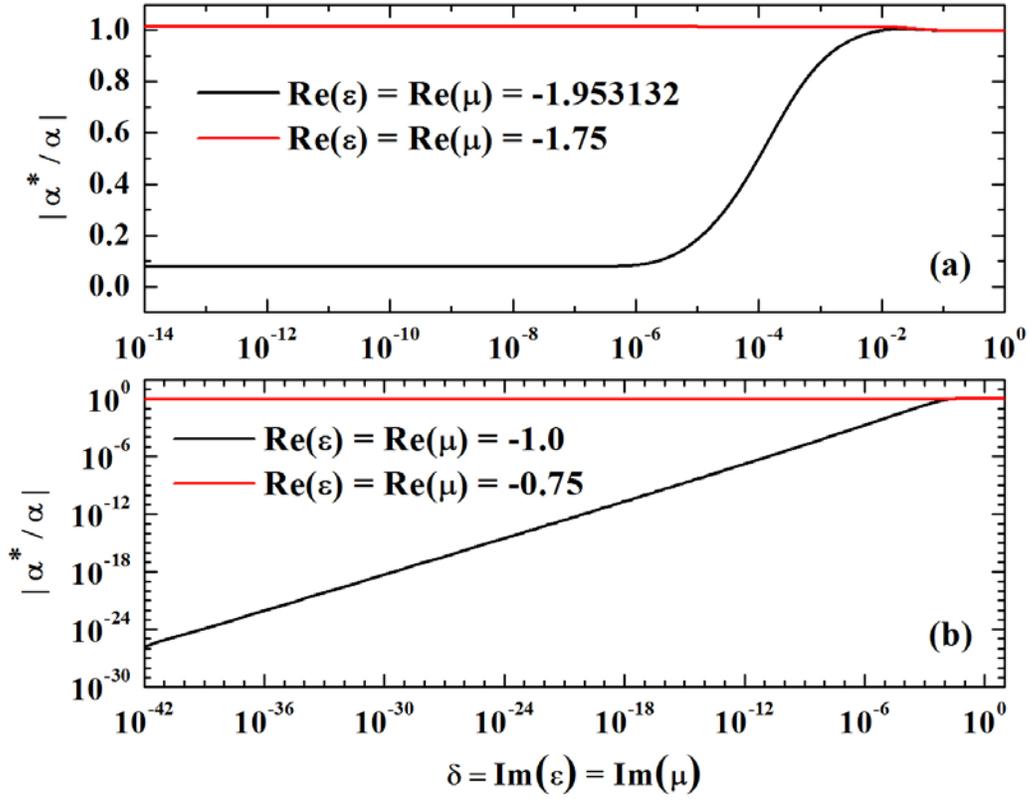

**FIG. 4. (Color online) The effective polarizability as a function of the imaginary part of the permittivity of the slab with different values of $\mathrm{Re}(\varepsilon) = \mathrm{Re}(\mu) = x$ and $z_d = d/5$. The imaging and cloaking properties of the metamaterial slab for some configurations are demonstrated in the Auxiliary Material.**

# Auxiliary material of "Metamaterial slab as a lens, a cloak or something in between"

Jian Wen Dong[1,2], Hui Huo Zheng[2], Yun Lai[2], He Zhou Wang[1], and C. T. Chan[2]

[1.] State Key Laboratory of Optoelectronic Materials and Technologies, Sun Yat-Sen (Zhongshan) University, Guangzhou 510275, China

[2.] Department of Physics, The Hong Kong University of Science and Technology, Hong Kong, China

**Part A. Asymptotic behaviors of the Green's function (for the source dipole located on the z-axis, with electric field polarized parallel to the slab)**

**A1. Asymptotic behaviors of $W_{yy}^{ref}$**

The reflection part of the dyadic Green's function is [Eq. (3) in the text],

$$W_{yy}^{ref} = \frac{i}{8\pi} \int_0^\infty \frac{k_{//}}{k_{0z}} dk_{//} \left( R^{TE} - \frac{k_{0z}^2}{k_0^2} R^{TM} \right) e^{2ik_{0z}z_d} . \tag{A1}$$

Here, we give the asymptotic behaviors of $W_{yy}^{ref}$ in the following limits: (i) $\varepsilon = \mu = -1 + i\delta, \delta \to 0$, (ii) $\varepsilon, \mu \neq -1, z_d \to 0$, (iii) $\varepsilon = -1 + i\delta, \mu \neq -1, \delta \to 0$, and (iv) $\mu = -1 + i\delta, \varepsilon \neq -1, \delta \to 0$.

**(i) The limit $\varepsilon = \mu = -1 + i\delta, \delta \to 0$ [Derivation of Eq. (5) in the text]**

Consider the reflection coming from each $k_{//}$ component. For $k_{//} \leq k_0$, we have $|R(k_{//})| \leq 2 \left| \frac{1-\zeta^2}{(1-\zeta)^2} \right|$ which implies $\lim_{\delta \to 0} |R(k_{//})| = 0$. This implies that the cloaking

effect in the perfect lens has nothing to do with the propagating components. For $k_{//} > k_0$, if $\delta$ is small, we have $1 - \zeta^2 = -2i\delta \frac{k_{//}^2}{k_{//}^2 - k_0^2} + O(\delta^2)$ and $1 + \zeta = -i\delta \frac{k_{//}^2}{k_{//}^2 - k_0^2} + O(\delta^2)$. So if $k_{//} \gg k_0$ (e. g. $k_{//} = 10k_0$) we have $\zeta \cong -1 - i\delta$, and $R^{TE} = R^{TM} \sim \frac{2i\delta e^{\kappa d}}{\delta^2 e^{\kappa d} + 4e^{-\kappa d}}$, where $\kappa = \sqrt{k_{//}^2 - \varepsilon \mu k_0^2}$. For any fixed $\delta$, we find when $k_{//} > (1/d)\log(2/\delta)$, $R(k_{//}) \to 2i/\delta$. Hence, for sufficiently large $k_{//}$, the reflection of the evanescent wave is very strong, and when we also take into account of the factor $e^{2ik_{0z}z_d}$ which is exponentially decaying for evanescent components, the reflection mostly comes from large $k_{//}$ Fourier components with its optimal value at $k_{//} \sim |\log \delta|/d$. It is straightforward to show that the reflection due to evanescent waves can be approximated as

$$\begin{aligned} W_{yy}^{ref} &\sim \frac{1}{4\pi} \int_{k_0}^{\infty} d\kappa \left[ \frac{i\delta e^{\kappa d\gamma}}{\delta^2 e^{\kappa d} + 4e^{-\kappa d}} + \frac{\kappa^2}{k_0^2} \frac{i\delta e^{\kappa d\gamma}}{\delta^2 e^{\kappa d} + 4e^{-\kappa d}} \right] \\ &= \frac{i}{4\pi d \delta^\gamma} \int_{\delta e^{k_0 d}}^{\infty} dx \frac{x^\gamma}{x^2 + 4} \left[ 1 + \frac{(\log x - \log \delta)^2}{k_0^2 d^2} \right] \end{aligned}. \tag{A2}$$

Here, $\gamma = 1 - 2z_d/d$ and $x = \delta e^{\kappa d}$. Hence, as $\delta \to 0$, it yields Eq. (5) in the text,

$$W_{yy}^{ref} \to \frac{i}{4\pi d \delta^\gamma} \left[ I_0(\gamma) \left( 1 + \frac{\log^2 \delta}{k_0^2 d^2} \right) - 2 \frac{\log \delta}{k_0^2 d^2} I_1(\gamma) + \frac{I_2(\gamma)}{k_0^2 d^2} \right], \tag{A3}$$

where $I_n(\gamma) = \int_0^{\infty} \frac{x^\gamma \log^n x}{x^2 + 4} dx$. It is clearly seen that $\lim_{\delta \to 0} W_{yy}^{ref} \to \infty$ if $z_d < d/2$ and $\lim_{\delta \to 0} W_{yy}^{ref} \to 0$ if $z_d > d/2$. It indicates that the Veselago slab with $\varepsilon = \mu = -1 + i\delta$ have cloaking effect with a critical distance $d/2$ in the zero absorption limit. This was noted by Milton and Nicorovici for two-dimensional line dipole configurations [7].

**(ii) The limit $\varepsilon, \mu \neq -1, z_d \to 0$ [Derivation of Eq. (4) in the text]**

As $z_d$ approaches zero, the evanescent component will dominate. The reflection coefficients approach to $R^{TE} \to \dfrac{(\mu-1)}{\mu+1}$ if $k_{//} \gg \dfrac{1}{d}\log\left|\dfrac{\mu-1}{\mu+1}\right|$. Here, we define $\kappa_0 = C\max\left\{\dfrac{1}{d}\log\left|\dfrac{\mu-1}{\mu+1}\right|, \dfrac{1}{d}\log\left|\dfrac{\varepsilon-1}{\varepsilon+1}\right|, \sqrt{\varepsilon\mu}k_0\right\}$, where the prefactor $C$ is a constant much bigger than 1 (10 would be a very safe choice). We consider those evanescent components with $k_{//} > \kappa_0$, in which $k_{0z} \sim ik_{//}$, $k_{//} \equiv \kappa$, $R^{TE} \sim (\mu-1)/(\mu+1) \equiv R^{TE}_{\lim}$ and $R^{TM} \sim (\varepsilon-1)/(\varepsilon+1) \equiv R^{TM}_{\lim}$. Hence, it yields Eq. (4) in the text,

$$W^{ref}_{yy} \to \frac{1}{8\pi}\int_{\kappa_0}^{\infty} d\kappa\, e^{-2\kappa z_d}\left[R^{TE}_{\lim} + \frac{\kappa^2}{k_0^2}R^{TM}_{\lim}\right] = \frac{1}{16\pi}\left[\frac{R^{TE}_{\lim}}{z_d} + R^{TM}_{\lim}\left(\frac{\kappa_0^2}{k_0^2 z_d} + \frac{\kappa_0}{k_0^2 z_d^2} + \frac{1}{2k_0^2 z_d^3}\right)\right]e^{-2\kappa_0 z_d}. \quad \text{(A4)}$$

Therefore, as $z_d \to 0$, $W^{ref}_{yy}$ diverges as $e^{-2\kappa_0 z_d}/z_d^3$. It indicates that $\alpha^*$ goes to zero smoothly with a asymptotic form $\sim z_d^3$, and a slab with $\varepsilon, \mu \neq -1$ has no critical-distance cloaking effect.

**(iii) The limit $\varepsilon = -1+i\delta, \mu \neq -1, \delta \to 0$ [Defining the "quasistatic limit" in the text]**

In this case, we will focus on the reflection coefficient for TM mode. We have $\zeta = \dfrac{\sqrt{\varepsilon\mu k_0^2 - k_{//}^2}}{\varepsilon\sqrt{k_0^2 - k_{//}^2}}$ and $\zeta^2 - 1 \cong 2i\delta + \dfrac{(\mu+1)k_0^2}{k_{//}^2}$, which lead to the following approximate expression of the reflection coefficient for large $k_{//} (\equiv \kappa)$,

$$R^{TM}(\kappa) = \frac{-(1-\zeta^2)(e^{ik_z d} - e^{-ik_z d})}{(1+\zeta)^2 e^{-ik_z d} - (1-\zeta)^2 e^{ik_z d}} \cong -\frac{\left[2i\delta + \frac{(1+\mu)k_0^2}{\kappa^2}\right] e^{\kappa d}}{\left[i\delta + \frac{(1+\mu)k_0^2}{2\kappa^2}\right]^2 e^{\kappa d} - 4e^{-\kappa d}}. \quad (A5)$$

We first consider the limit $k_0 \to 0$. In this limit, Eq. (A5) becomes $R^{TM}(\kappa) = \frac{2i\delta e^{\kappa d}}{\delta^2 e^{\kappa d} + 4e^{-\kappa d}}$, which is the same as a Veselago slab with absorption $\delta$ in finite frequency regime. Hence, we expect that the integral in $W_{yy}^{ref}$ diverges as $\delta \to 0$ for $z_d < d/2$. However, in this case, since $k_0 \to 0$, it is better for us to examine the quantity $4\pi k_0^2 W_{yy}^{ref}$ (see the expression of $\alpha^*$),

$$\lim_{k_0 \to 0} 4\pi k_0^2 W_{yy}^{ref} \to \frac{i}{d^3 \delta^\gamma}\left[I_0(\gamma)\log^2 \delta - 2I_1(\gamma)\log \delta + I_2(\gamma)\right]. \quad (A6)$$

We find that if $z_d < d/2$, $\lim_{k_0 \to 0} 4\pi k_0^2 W_{yy}^{ref}$ diverges as $\delta \to 0$, with the asymptotic form $C\delta^{-\gamma}\log^2 \delta$. This indicates that a slab with $\varepsilon = -1+i\delta, \mu \neq -1$ can have finite distance cloaking effect in the quasistatic limit, with a critical distance $d/2$.

From Eq. (A5), we know if

$$k_0 < \sqrt{2}(1/d)\delta^{1/2}\log(2/\delta)/\sqrt{1+\mu}, \quad (A7)$$

we have $R^{TM} \cong \frac{2i\delta e^{\kappa d}}{\delta^2 e^{\kappa d} + 4e^{-\kappa d}}$ for all $k_\parallel > (1/d)\log(2/\delta)$. In this case, a slab with $\varepsilon = -1+i\delta, \mu \neq -1$ can have cloaking effect similar to a Veselago slab, i.e., cloaking with a critical distance $d/2$. Eq. (A7) shows the upper bound of the working wavelength, and gives the quantitative definition of "quasistatic limit". Note that the upper bound is related to the slab thickness $d$ and $\mu$. In the perspective of Eq. (A7), the Veselago slab with $\mu \to -1$ corresponds to infinite upper bound, so that the Veselago lens has critical cloaking effect for arbitrary frequency.

We now consider the finite frequency regime and take $\delta \to 0$. We have

$$R^{TM}(\kappa) = -\frac{\frac{(1+\mu)k_0^2}{\kappa^2}e^{\kappa d}}{\left[\frac{(1+\mu)k_0^2}{2\kappa^2}\right]^2 e^{\kappa d} - 4e^{-\kappa d}}.$$ If $z_d \to 0$, the integral related to TM mode

in the $4\pi k_0^2 W_{yy}^{ref}$ can be approximated as,

$$f(z_d) \sim -\frac{1}{16(1+\mu)k_0^2 z_d^5}\left[(2\tilde{\kappa}z_d)^4 + 4(2\tilde{\kappa}z_d)^3 + 12(2\tilde{\kappa}z_d)^2 + 24(2\tilde{\kappa}z_d) + 24\right]e^{-2\tilde{\kappa}z_d}$$
$$\sim \frac{3}{2(1+\mu)k_0^2 z_d^5}e^{-2\tilde{\kappa}z_d}$$
(A8)

Here $\tilde{\kappa}$ is a sufficient large value such that $\left[\frac{(1+\mu)k_0^2}{2\tilde{\kappa}^2}\right]^2 e^{\tilde{\kappa}d} \gg 4e^{-\tilde{\kappa}d}$. In fact, due to the exponential factor, $\left[\frac{(1+\mu)k_0^2}{2\kappa^2}\right]^2 e^{\kappa d}$ will very quickly dominate as $\kappa$ increase, therefore, $\tilde{\kappa}$ in general need not to be very large. For example, if $\mu = -2$,

$k_0 d = 1$ , $\tilde{\kappa} = 7/d$ , $\left[\frac{(1+\mu)k_0^2}{2\tilde{\kappa}^2}\right]^2 e^{\tilde{\kappa}d} \sim 40 e^{-\tilde{\kappa}d}$ . Here

$$f(z_d) = -\frac{1}{2}k_0^2 \int_{k_0}^{\tilde{\kappa}} \frac{(1+\mu)e^{\kappa(d-2z_d)}}{\left[\frac{(1+\mu)k_0^2}{2\kappa^2}\right]^2 e^{\kappa d} - 4e^{-\kappa d}} d\kappa \;,$$ which is finite when $z_d \to 0$

($|f(z_d)| < |f(0)|$). Hence, we have $W_{yy}^{ref} \sim C/z_d^5$ as $z_d \to 0$. Therefore, in the finite

frequency regime beyond the quasi-static regime, a slab with $\varepsilon = -1 + i\delta, \mu \neq -1$

does not have critical-distance cloaking effect, but the suppression effect is much more stronger than $\varepsilon, \mu \neq -1$.

**(iv) The limit** $\mu = -1 + i\delta, \varepsilon \neq -1, \delta \to 0$

First we consider the limit $k_0 \to 0$. Similar to case (iii), when $z_d < d/2$ and $\delta \to 0$, $W_{yy}^{ref}$ will diverge with the asymptotic form of $W_{yy}^{ref} \to \dfrac{i}{4\pi d \delta^{\gamma}} I_0(\gamma) + \dfrac{1}{16\pi} R_{\lim}^{TM} \left( \dfrac{\kappa_0^2}{k_0^2 z_d} + \dfrac{\kappa_0}{k_0^2 z_d^2} + \dfrac{1}{2 k_0^2 z_d^3} \right) e^{-2\kappa_0 z_d}$. However, $4\pi k_0^2 W_{yy}^{ref}$ does not diverge if $k_0 \to 0$. If $k_0$ satisfies the quasistatic limit condition $k_0 < \sqrt{2}(1/d)\delta^{1/2} \log(2/\delta)/\sqrt{1+\varepsilon}$, we have

$$4\pi k_0^2 W_{yy}^{ref} \to \dfrac{2i\delta^{2z_d/d}}{(1+\varepsilon)d^3} I_0(\gamma) \log^2(2/\delta) + \dfrac{1}{4} R_{\lim}^{TM} \left( \dfrac{\kappa_0^2}{z_d} + \dfrac{\kappa_0}{z_d^2} + \dfrac{1}{2z_d^3} \right) e^{-2\kappa_0 z_d}, \quad (A9)$$

The first term of Eq. (A9) goes to zero as $\delta \to 0$, and hence $4\pi k_0^2 W_{yy}^{ref}$ remain finite (equals to the second term) if $z_d \neq 0$ when $\delta \to 0$, i.e., $\alpha^*$ goes to zero smoothly as $z_d \to 0$ (the asymptotic behaviors is $\sim C z_d^3$). Therefore, there is no critical distance for $\mu = -1 + i\delta, \varepsilon \neq -1$ in quasistatic limit.

Now consider the finite frequency and $\delta \to 0$. If $z_d \to 0$, the integral related to TE mode in the $4\pi k_0^2 W_{yy}^{ref}$ can be approximated as

$g(z_d) - \dfrac{e^{-2\tilde{\kappa} z_d}}{4(1+\varepsilon)z_d^3} \left[ (2\tilde{\kappa} z_d)^2 + 2(2\tilde{\kappa} z_d) + 2 \right] \sim \dfrac{-ie^{-2\tilde{\kappa} z_d}}{2(1+\varepsilon)z_d^3}$, where $\tilde{\kappa}$ has the same definition as Eq. (A8), and $g(z_d) = -\dfrac{1}{2} k_0^2 \int_{k_0}^{\tilde{\kappa}} \dfrac{\dfrac{(1+\varepsilon)k_0^2}{\kappa^2} e^{\kappa(d-2z_d)}}{\left[ \dfrac{(1+\varepsilon)k_0^2}{2\kappa^2} \right]^2 e^{\kappa d} - 4e^{-\kappa d}} d\kappa$, which is finite when $z_d \to 0$ ($|g(z_d)| < |g(0)|$). Hence, we note that the asymptotic behavior is $4\pi k_0^2 W_{yy}^{ref} \sim C/z_d^3$. Therefore, a slab with $\mu = -1 + i\delta, \varepsilon \neq -1$ does not have critical-distance cloaking effect.

## A2. Asymptotic behaviors of $W_{yy,img}^{tran}$ ($\varepsilon = \mu = -1+\delta, \delta \to 0$)

The transmission part of the dyadic Green's function in the image point is [12],

$$4\pi k_0^2 W_{yy,img}^{tran} = \frac{ik_0^2}{2} \int_0^\infty \frac{k_{//}}{k_{0z}} dk_{//} \left( T^{TE} + \frac{k_{0z}^2}{k_0^2} T^{TM} \right) e^{i2k_{0z}d}, \tag{A10}$$

where $T = \dfrac{4\zeta e^{-ik_z d} e^{-ik_{0z} d}}{(\zeta+1)^2 e^{-i2k_z d} - (\zeta-1)^2}$. For $k_{//} \leq k_0$, we have

$\lim_{\delta \to 0} \left| T(k_{//}) e^{2ik_{0z}d} \right| = \lim_{\delta \to 0} 4 \left| \dfrac{\zeta}{(1-\zeta)^2} \right| = 1$. The propagating modes should have a transfer function of $\sim 1$, but as $\alpha^* \to 0$, the propagating modes cannot provide any intensity for image formation. The evanescent modes with $k_{//} > k_0$ is more subtle. If $\delta$ is small, we have $T^{TE} = T^{TM} \sim \dfrac{4e^{\kappa d}}{\delta^2 e^{\kappa d} + 4e^{-\kappa d}}$, where $\kappa = \sqrt{k_{//}^2 - \varepsilon\mu k_0^2}$. For any fixed $\delta$, we find when $k_{//} > (1/d)\log(2/\delta)$, $T(k_{//}) \to 4/\delta^2$. As $T(k_{//})$ is capped by $4/\delta^2$, $T(k_{//})e^{2ik_{0z}d}$ actually drops to zero for sufficiently big values $k_{//}$ ($k_{//} > (1/d)\log(2/\delta)$) due to the exponential factor $e^{2ik_{0z}d}$. We note that $T(k_{//})e^{2ik_{0z}d} = 1$ for all $k_{//}$ if simply put $\delta = 0$ in the calculations without considering the limit carefully. Mathematically, the transmission due to evanescent waves can be approximated as $2k_0^2 \dfrac{1}{d} \left[ I_0' \left( 1 + \dfrac{\log^2 \delta}{k_0^2 d^2} \right) - 2 \dfrac{\log \delta}{k_0^2 d^2} I_1' + \dfrac{I_2'}{k_0^2 d^2} \right]$, where

$I_n' = \int_0^\infty \dfrac{\log^n x}{x(x^2+4)} dx$. We note that when $z_d < \dfrac{d}{2}$, $4\pi k_0^2 W_{yy}^{ref}$ is divergent as $\delta \to 0$ and it is more divergent than $4\pi k_0^2 W_{yy,img}^{tran}$. As $\alpha^* \sim (4\pi k_0^2 W_{yy}^{ref})^{-1}$, we find $\lim_{\delta \to 0} \alpha^* 4\pi k_0^2 W_{yy,img}^{tran} \to 0$. It means that evanescent waves also cannot contribute to any intensity at the image point. It indicates that in the pursuit of an ideal superlens, the image will approach perfect resolution but it has zero brightness.

**Part B. Asymptotic behaviors of the Green's function if the electric field of the source dipole is polarized perpendicular to the slab**

**B1. Asymptotic behaviors of $W_{zz}^{ref}$**

In this section, we will discuss the cloaking effect if the source dipole is polarized perpendicular to the slab. This corresponds to the principal value of effective polarizability tensor, $\alpha_{zz}^*$, while Part A corresponds to $\alpha_{yy}^* = \alpha_{xx}^*$. We will see that the results in Part B are qualitatively the same as the case in Part A. In this section, we should consider the excitation of the perpendicular component of the Green's function, which is given by

$$4\pi k_0^2 W_{zz}^{ref} = i\int_0^\infty \frac{k_{//}^3}{k_{0z}} dk_{//} R^{TM} e^{2ik_{0z}z_d} \ . \tag{B1}$$

**(i) The limit $\varepsilon = -1 + i\delta$, $\mu = -1$**

In this case, we have $\lim_{\delta \to 0} 4\pi k_0^2 W_{zz}^{ref} \to \frac{2i}{d^3 \delta^\gamma} \left[ I_0(\gamma) \log^2 \delta - 2I_1(\gamma) \log \delta + I_2(\gamma) \right]$. We find that if $z_d < d/2$, $\lim_{\delta \to 0} 4\pi k_0^2 W_{zz}^{ref}$ diverges as $\delta \to 0$, with the asymptotic form $C\delta^{-\gamma} \log^2 \delta$. This indicates that it can have finite distance cloaking effect with a critical distance $d/2$.

**(ii) The limit $\varepsilon = -1 + i\delta$, $\mu \neq -1$**

We first consider the limit $k_0 \to 0$. In this limit, $4\pi k_0^2 W_{zz}^{ref}$ has the same asymptotic form as that of $\mu = -1$ (see section B1). So a slab with $\varepsilon = -1 + i\delta, \mu \neq -1$ can have finite distance cloaking effect in the quasistatic limit, with a critical distance $d/2$. Now we turn to consider finite frequency, zero absorption limit, we should have the asymptotic form of $R^{TM}(\kappa) = -\dfrac{\dfrac{(1+\mu)k_0^2}{\kappa^2}e^{\kappa d}}{\left[\dfrac{(1+\mu)k_0^2}{2\kappa^2}\right]^2 e^{\kappa d} - 4e^{-\kappa d}}$. If $z_d \to 0$, $4\pi k_0^2 W_{zz}^{ref}$ have the following asymptotic form,

$$2f(z_d) - \frac{1}{8(1+\mu)k_0^2 z_d^5}\left[(2\tilde{\kappa}z_d)^4 + 4(2\tilde{\kappa}z_d)^3 + 12(2\tilde{\kappa}z_d)^2 + 24(2\tilde{\kappa}z_d) + 24\right]e^{-2\tilde{\kappa}z_d}$$
$$\sim \frac{3}{(1+\mu)k_0^2 z_d^5}e^{-2\tilde{\kappa}z_d}$$
(B2)

where the definitions of $f(z_d)$ and $\tilde{\kappa}$ are the same on page 5. Hence, we have $W_{zz}^{ref} \sim C/z_d^5$ as $z_d \to 0$. Therefore, in the frequency regime, a slab with $\varepsilon = -1 + i\delta, \mu \neq -1$ does not have a critical cloaking distance.

**(iii) The limit $\varepsilon \neq -1$**

In this case, we have, $4\pi k_0^2 W_{zz}^{ref} \to \dfrac{1}{8z_d^3} R_{\lim}^{TM}\left[(2\kappa_0 z_d)^2 + 2(2\kappa_0 z_d) + 2\right]e^{-2\kappa_0 z_d} \sim C/z_d^3$.

Here we define $\kappa_0 = C \max\left\{(1/d)\log\left|(\varepsilon-1)/(\varepsilon+1)\right|, \sqrt{\varepsilon\mu}k_0\right\}$, where the prefactor C is a constant much bigger than 1. Therefore, a slab with $\varepsilon \neq -1$ does not have critical-distance cloaking effect.

**B2. Asymptotic behaviors of $W_{zz,img}^{tran}$ ($\varepsilon = \mu = -1+\delta, \delta \to 0$)**

In this case, we employ the similar procedures as those in section A2. As $\delta \to 0$, the asymptotic behavior of $W_{zz,img}^{tran}$ is

$$4\pi k_0^2 W_{zz,img}^{tran} = i\int_0^\infty \frac{k_{//}^3}{k_{0z}}dk_{//}T^{TM}e^{2ik_0 z d} \sim \frac{4}{d^3}\left(I_0'\log^2\delta - 2I_1'\log\delta + I_2'\right).$$ (B3)

As $4\pi k_0^2 W_{zz}^{ref}$ is more divergent than $4\pi k_0^2 W_{zz,img}^{tran}$ as $\delta \to 0$ when the dipole is placed with $d/2$ of the slab, and as $\alpha^* \sim (4\pi k_0^2 W_{zz}^{ref})^{-1}$, we find that $\lim_{\delta \to 0} \alpha^* 4\pi k_0^2 W_{zz,img}^{tran} \to 0$. The image of the dipole will have has zero brightness in the limit of no absorption in the lens.

**Part C: Cloaking effect for "folded geometry" slab**

In a "folded geometry" slab with $\varepsilon = \mu = diag(-\beta, -\beta, -1/\beta)$, $\beta > 0$, we have $R = \dfrac{-\left[1 - \zeta^2\right]\left(e^{ik_z d} - e^{-ik_z d}\right)}{(1+\zeta)^2 e^{-ik_z d} - (1-\zeta)^2 e^{ik_z d}}$, where $\zeta = \dfrac{\sqrt{\mu_x \varepsilon_y \omega^2 / c^2 - k_x^2 \mu_x / \mu_z}}{\mu_x \sqrt{\omega^2 / c^2 - k_x^2}}$ for TE mode, and $\zeta = \dfrac{\sqrt{\mu_y \varepsilon_x \omega^2 / c^2 - k_x^2 \varepsilon_x / \varepsilon_z}}{\varepsilon_x \sqrt{\omega^2 / c^2 - k_x^2}}$ for TM mode. For large $k_x (\equiv \kappa)$, we have $k_z \sim i\beta\kappa$, and $R^{TE} = R^{TM} \sim \dfrac{2i\delta\left[e^{\beta\kappa d} - e^{-\beta\kappa d}\right]}{\delta^2 e^{\beta\kappa d} + 4e^{-\beta\kappa d}}$, which is the same as a Veselago lens of thickness $\beta d$. Hence, we can expect that the cloaking effect also occurs in this kind of slab with a critical distance $\beta d / 2$. Here we numerically demonstrate the cloaking effect (for a 2D system) calculated by COMSOL package. The cylindrical object ($r = 0.005d, \varepsilon = 480$) is placed at a distance $z_d = d/2$ and $z_d = 5d/4$ in front of an anisotropic slab with material parameters $\varepsilon_y = -2, \mu_z = -0.5, \mu_x = -2$, and illuminated by a Gaussian beam with $\lambda = 0.125d$ and the waist size of $0.125d$. We see that the object can be cloaked [Fig. C(a)] or imaged [Fig. C(b)] by the slab depending on whether is it within or outside $z_d = \beta d / 2$.

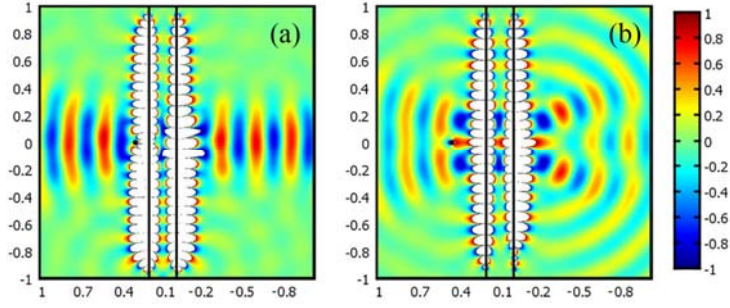

Fig. C. The "dipole" object (black dot) is (a) cloaked or (b) imaged by an anisotropic "folded geometry" slab.

**Part D: Numerical demonstrations of the cloaking and imaging properties**

In Fig. 4 in the text, we plotted $\left|\alpha^*/\alpha\right|$ for some values of $\varepsilon = \mu = x + i\delta$ to highlight the cloaking vs imaging effect. Here we show the field distributions for these configurations with a value of $\delta = 10^{-7}$. Using the Green's function method, the electric field at $\mathbf{R}$ can be written as $E_y(\mathbf{R}) = \omega^2 \mu_0 \left[ W_{yy,\mathbf{R}}^{tot} + \alpha^* 4\pi k_0^2 W_{yy,\mathbf{R}}^{tran} W_{yy,dipole}^{tot} \right] p_{src}$, where $W_{yy,\mathbf{R}}^{tot}$ ($W_{yy,dipole}^{tot}$) represents the yy-element of the total Green's function radiated at $\mathbf{R}$ (the passive dipole) by the external source, and $W_{yy,\mathbf{R}}^{tran}$ represents the Green's function radiated at $\mathbf{R}$ by the dipole object [12]. The color stands for $\log(|E_{tot}|)$. The source ($p_{src}$) is placed at *3d* to the left of the slab.

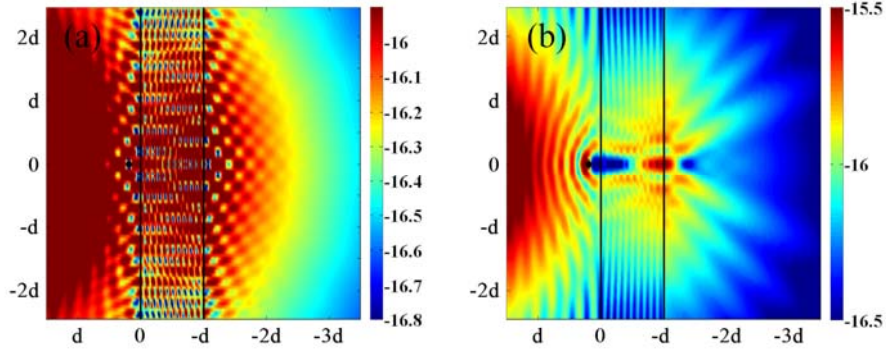

Fig. D1. (a) Partial cloaking and (b) imaging/scattering effect when the "dipole" object (black dot) is at a distance $z_d = d/5$ in front of the metamaterial slab (between solid lines). The left and right panel corresponds to the black and red curves respectively in Fig. 4(a).

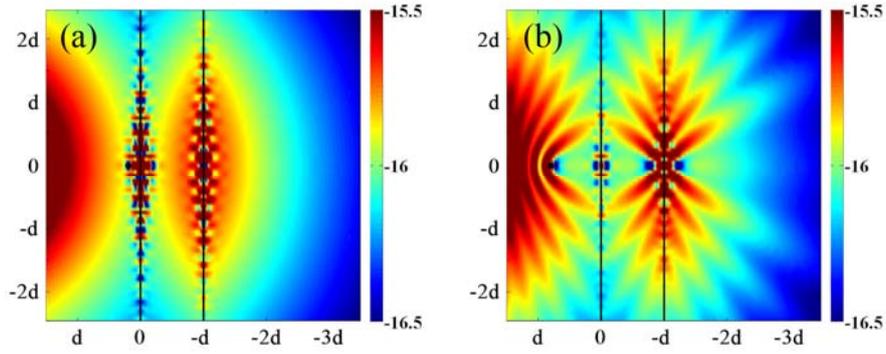

Fig. D2. (a) Cloaking and (b) lensing effect of the lossy Veselago slab ($\varepsilon = \mu = -1 + 10^{-7}i$) when the object locates at $z_d = d/5$ and $z_d = 4d/5$.